\documentclass[12pt]{article}
\usepackage[utf8]{inputenc}
\usepackage{amsmath}
\usepackage{amsfonts}
\usepackage{amssymb}
\usepackage{slashed}
\usepackage{graphics}
\usepackage[framemethod=TikZ]{mdframed}
\mdfdefinestyle{MyFrame}{%
    linecolor=blue,
    outerlinewidth=2pt,
    roundcorner=20pt,
    innertopmargin=\baselineskip,
    innerbottommargin=\baselineskip,
    innerrightmargin=20pt,
    innerleftmargin=20pt,
    backgroundcolor=white!50!white }
\begin{document}
\title{On Generating a Lagrangian for Higher Dimensional Gravity}
\author{Theo Verwimp{$^{\dagger}$}\\
e-mail: theo.verwimp@telenet.be}
\renewcommand{\today}{Februari 10, 2021}
\maketitle
The Lovelock Lagrangian is for even dimension $D$ obtained from Weil polynomials on the Lie algebra of the Lorentz group $SO(1,D-1)$. The procedure for generating it is related to the Weil homomorphism that converts Lie algebra invariants into cohomology classes.\\
\section{Introduction}
If $(M,g)$ is a $D$-dimensional spacetime, then the gravitational action is given by the integral of a $D$-form, a Lagrangian on $M$. A particularly interesting gravity Lagrangian, which is a generalization to more than 4 dimensions of the Einstein Lagrangian because it yields divergence-free symmetric second-order field equations, is the Lovelock Lagrangian[1]. In terms of the curvature 2-form $\Omega(\omega)$ computed from a connection 1-form $\omega$ compatible with the metric $g$, and the orthonormal coframe $\theta^{i}$ defined by $g=\eta_{ij}\theta^{i}\otimes \theta^{j}$, $\eta_{ij}$ the Minkowski metric, it takes the form[2]:
\begin{equation}
\mathcal{L}=\sum_{p=0}^{[D/2]-1} \frac{\lambda_{p}}{(D-p)!} \varepsilon _{i_{1}\cdot\cdot\cdot i_{D}} \Omega ^{i_{1}i_{2}} \wedge \cdot\cdot\cdot \wedge \Omega ^{i_{2p-1}i_{2p}} \wedge \theta^{i_{2p+1}} \wedge \cdot\cdot\cdot \wedge \theta^{i_{D}}
\end{equation}

where $\varepsilon _{i_{1}...i_{D}}$ is the totally antisymmetric tensor with $\varepsilon_{1...D}=1$. The Lagrangian is a linear combination of terms associated with all even dimensions below $D$, and in which each term is obtained by the continuation to dimension $D$ of the Euler form from a dimension lower than $D$. If $D$ is even, a term with $p=D/2$ can be added which is proportional to the Euler form in dimension $D$. Its integral over a compact submanifold is a topological invariant and gives no contribution to the field equations in the classical theory[3]. For these reasons, the theory following from this Lagrangian has been called a dimensionally continued topological gravitation theory[4]. The method used here to generate the Lagrangian (1) is a generalisation to higher dimension (but for zero torsion), of the work done by Kakazu and Matsumoto in ref.[5].

†Former affiliated with: Physics Department; U.I.A., Universiteit Antwerpen Belgium

†On retirement from ENGIE Laborelec, Belgium 

\pagebreak  

\section{A Lagrangian form on a principal fibre bundle}
Let $G$ be a Lie group with Lie algebra $\mathcal{G}$ and $\lbrace \xi\rbrace _{i=1\cdot\cdot\cdot r}$ a basis for the dual space $\mathcal{G}^{\ast}$. Let $P_{G}(\mathcal{G})$ denote the algebra of homogeneous $Ad(G)$-invariant polynomial functions on $\mathcal{G}$. This algebra may be identified with the algebra $S_G(\mathcal{G})$ of symmetric $Ad(G)$-invariant multilinear mappings on $\mathcal{G}$[6,7]. The identifying isomorphism $\mathcal{S}:P_{G}(\mathcal{G})\rightarrow S_{G}(\mathcal{G})$ is given by
\begin{equation}
(\mathcal{S}f)(X_{1}, \cdot\cdot\cdot\, ,X_{k})=a_{i_{1}\cdot\cdot\cdot i_{k}}\xi^{i_{1}}(X_{1})\cdot\,\cdot\cdot\cdot\, \cdot \xi^{i_{k}}(X_{k}),\;\; X_{1},\cdot\cdot\cdot ,X_{k}\in \mathcal{G}
\end{equation}
 where $a_{i_{1}\cdot\cdot\cdot i_{k}}\xi^{i_{1}}\cdot\,\cdot\cdot\cdot\, \cdot  \xi^{i_{k}}$ is the unique expression of the polynomial function $f$ of degree $k$ in the basis $\lbrace \xi^{i}\rbrace$ for $\mathcal{G^{\ast}}$.\\
Now, let $\pi=P\rightarrow M$ be a principal fibre bundle over a manifold $M$ (of dimension $D$) and with structure group $G$. If $\alpha_{1}, \cdot\cdot\cdot\, , \alpha_{k}$ are $\mathcal{G}$-valued forms on $P$ of degree $q_{1}, \cdot\cdot\cdot\, , q_{k}$, respectively, the we define for every 'Weil polynomial' $F\in S_{G}(\mathcal{G})$ an $\mathbb{R}$-valued 'Weil-form' $F(\alpha_{1}, \cdot\cdot\cdot\, , \alpha_{k})$ of degree $q=\sum_{i=1}^{k} q_{i}$ on $P$ by[6]:
\begin{multline}
F(\alpha_{1}, \cdot\cdot\cdot\, , \alpha_{k})(X_{1}, \cdot\cdot\cdot\, ,X_{q})= \\
\frac{1}{q_{1}!\cdot\cdot\cdot q_{k}!}\sum_{\sigma =S_{q}}\varepsilon (\sigma)F(\alpha_{1}(X_{\sigma(1)},\cdot\cdot\cdot ,X_{\sigma (q_{1})}), \cdot\cdot\cdot\, , \alpha_{k}(X_{\sigma(\sum_{1}^{k-1}q_{i}+1)},\cdot\cdot\cdot ,X_{\sigma (q)}))
\end{multline}
for $X_{1}, \cdot\cdot\cdot\, ,X_{q}\in T_{u}(P)$(tangent space of $P$ at $u$), where the summation is taken over all permutations $\sigma$ of $(1,\cdot\cdot\cdot ,q)$ and $\varepsilon (\sigma)$ is the sign of the permutation. If$\lbrace E_{i}\rbrace_{i=1,\cdot\cdot\cdot,r}$ is a basis for $\mathcal{G}$, such that $\alpha_{i}=\alpha_{i}^{j}E_{j}$ ($\alpha_{i}^{j}\in \Lambda ^{q_{i}}(P,\mathbb{R}) $ the space of $\mathbb{R}$-valued $q_{i}$-forms on P), then one obtains from the multilinearity of $F$ and the definition of the wedge-product, the result:
\begin{equation}
F(\alpha_{1}, \cdot\cdot\cdot\, , \alpha_{k})=F(E_{j_{1}},\cdot\cdot\cdot ,E_{j_{k}})\alpha_{1}^{j_{1}}\wedge \cdot\cdot\cdot \wedge \alpha_{k}^{j_{k}}.
\end{equation}\\
Our final objective is the construction of a Lagrangian form ($L$-form) on P such that it projects to $M$ uniquely and then identify this projection with a gravitational Lagrangian on $M$. The projection of the $L$-form will be unique under certain well defined conditions for $\alpha_{i}$[5,6]. If the $\alpha_{i}$ of degree $q_{i}$ are elements of $\bar{\Lambda}^{q_{i}}(P, \mathcal{G})$, that is, \\
(i) $\alpha_{i}(R_{g\ast}X_{1},\cdot\cdot\cdot ,R_{g\ast}X_{q_{i}})=Ad(g^{-1})\alpha_{i}(X_{1}, \cdot\cdot\cdot , X_{q_{i}})$ for right translations $R_{g}, g\in \mathcal{G}$,\\
(ii) $\alpha_{i}(X_{1}, \cdot\cdot\cdot , X_{q_{i}})=0$ whenever at least one of the $X_{i}\in T_{u}(P)$ is vertical, then the Weil-form $F(\alpha_{1}, \cdot\cdot\cdot\, , \alpha_{k})$ on $P(F\in S_{G}(\mathcal{G}))$ of degree $q=\sum_{i=1}^{k} q_{i}$ will project to a unique $q$-form, $\bar{F}(\alpha_{1}, \cdot\cdot\cdot\, , \alpha_{k})$ on $M$ , i.e.,\\
\begin{equation}
F(\alpha_{1}, \cdot\cdot\cdot\, , \alpha_{k})=\pi^{\ast}(\bar{F}(\alpha_{1}, \cdot\cdot\cdot\, , \alpha_{k}))
\end{equation}\\
Now, consider on $P$ the $D=2m$-form $L_{m}\in S_{G}\medskip^{m}(\mathcal{G})$ formed by summation over algebraically independent Weil Polynomials of degree m on the Lie-algebra $\mathcal{G}$:
\begin{multline}
L_{m}(\Omega (\omega)+\alpha h(\theta))\equiv L_{m}(\Omega (\omega)+\alpha h(\theta),\cdot\cdot\cdot , \Omega (\omega)+\alpha h(\theta))\\
=L_{m}(\Omega(\omega),\cdot\cdot\cdot ,\Omega(\omega))+m\alpha L_{m}(\Omega(\omega),\cdot\cdot\cdot ,\Omega(\omega), h(\theta))+\\
\cdot\cdot\cdot +m\alpha^{m-1}L_{m}(\Omega(\omega),h(\theta),\cdot\cdot\cdot ,h(\theta))+\alpha^{m}L_{m}(h(\theta),\cdot\cdot\cdot ,h(\theta)), 
\end{multline}
where $\alpha$ is a constant, $\Omega(\omega)=d\omega + \omega \wedge \omega \in \bar{\wedge}^{2}(P,\mathcal{G})$ the curvature $2$-form calculated from a connection $1$-form $\omega$ on $P$[7] and
\begin{equation}
h(\theta)=\theta^{i}\wedge \theta^{j}h(e_{i},e_{j})\in \bar{\wedge}^{2}(P,\mathcal{G})
\end{equation}
with $h:\mathbb{R}\times\mathbb{R}\rightarrow \mathcal{G}$ the multilinear antisymmetric mapping which satisfies[5]:\\ $h(A(g)e_{i}, A(g)e_{j})=Ad(g)h(e_{i},e_{j})$, for $g\in G$, $A(g)$ a $D$-dimensional representation of $g$ and $\lbrace e_{i} \rbrace $ a basis for $\mathbb{R}^{D}$. Further, $\theta = \theta^{i}e_{i}$ is an element of $\bar{\wedge}^{1}(P,\mathbb{R}^{D})$ and we will identify it with the $D$-bein form.[5]\\
The gauge invariance of the $D$-form $L_{m}(\Omega (\omega)+\alpha h(\theta))$ on $P$, that is, $L_{m}(\Omega (s^{\ast}\omega)+\alpha h(s^{\ast}\theta))=L_{m}(\Omega (\omega)+\alpha h(\theta))$ for each diffeomorphism $s:P\rightarrow P$ such that $s(ug)=s(u)g$ and $\pi (u)=\pi (s(u))$ for each $g \in G$ and $u\in P$[8], is a consequence of the $Ad(G)$-invariance of $L_{m}$ and the fact that $\Omega(\omega)$ as well as $h(\theta)$ are elements of $\bar{\wedge}^{2}(P,\mathcal{G})$. The proof is the same as given in Ref.[5].

\section{The Lovelock Lagrangian}

From now on let $P(M,G,\pi)\equiv F(M)$ be the orthonormal frame bundle over $D=2m$-dimensional spacetime $M$ with $G=SO(1,D-1)$. Generators of the Lie-algebra $\mathcal{G}=so(1,D-1)$ are in $D\times D$-matrix representation given by
\begin{equation}
(J_{ij})^{k}_{l}=\delta^{k}_{i}\eta_{jl}-\delta^{k}_{j}\eta_{il}\;\; \eta=(diag(-1,1,\cdot\cdot\cdot ,1).
\end{equation}
Since $J_{ij}=-J_{ji}$ ther are only $D/2)(D-1)$ independent generators. Therefore a basis for the Lie-algebra $\mathcal{G}$ is given by $\lbrace J_{ij}\rbrace_{i<j}$. In this basis we have
\begin{equation}
\Omega(\omega)=\frac{1}{2}\Omega^{ij}(\omega)J_{ij},\;\; h(\theta)=\frac{1}{2}\theta^{i}\wedge \theta^{j}J_{ij}
\end{equation}
The expression for $h(\theta)$ follows from (7) and $h(e_{k},e_{l})=\frac{1}{2}h^{ij}(e_{k},e_{l})J_{ij}$ where now $h^{ij}(e_{k},e_{l})=\frac{1}{2}(\delta^{i}_{k}\delta^{j}_{l}-\delta^{i}_{l}\delta^{j}_{k})$[5].  We also identify $\theta\in \bar{\wedge}^{1}(P,\mathbb{R}^{D})$as the canonical $1$-form or $D$-bein form.[7]
From the theory of Lie-algebras it is well known[6,7,10], that algebraic  independent $Ad(G)$-invariant polynomial functions on $\mathcal{G}$ are given by the characteristic coefficients $f_{k}(X)$ in the expension in powers of $\lambda$ of the $DXD$-matrix determinant:
\begin{equation}
det(\lambda+X)=\sum_{k=0}^{D}f_{k}(X)\lambda^{D-k} , \;\; X\in \mathcal{G}.
\end{equation}
If $\lbrace \chi^{i}_{j}\rbrace$ is a basis for the dual space $\mathcal{G}^{\ast}:\chi^{i}_{j}(X)=X^{i}_{j}$ where $X\in \mathcal{G}$, then the polynomial functions $f_{k}$ will be given by[9]
\begin{equation}
f_{k}(X)=\frac{1}{k!}\delta^{i_{1}\cdot\cdot\cdot i_{k}}_{j_{1}\cdot\cdot\cdot j_{k}}X^{j_{1}}_{i_{1}}\cdot\,\cdot\cdot\cdot\,\cdot X^{j_{k}}_{i_{k}}=\frac{1}{k!}\delta^{i_{1}\cdot\cdot\cdot i_{k}}_{j_{1}\cdot\cdot\cdot j_{k}}\chi^{j_{1}}_{i_{1}}(X)\cdot\,\cdot\cdot\cdot\,\cdot \chi^{j_{k}}_{i_{k}}(X)
\end{equation}
where
\begin{equation}
\delta^{i_{1}\cdot\cdot\cdot i_{k}}_{j_{1}\cdot\cdot\cdot j_{k}}=\sum_{\sigma\in S_{k}}\varepsilon (\sigma)\delta^{i_{1}}_{j\sigma(1)}\cdot\,\cdot\cdot\cdot\,\cdot \delta^{i_{k}}_{j\sigma(k)}=-\frac{1}{(D-k)!}\varepsilon_{j_{1}\cdot\cdot\cdot\cdot j_{k}l_{k+1}\cdot\cdot\cdot l_{D}}\varepsilon^{i_{1}\cdot\cdot\cdot\cdot i_{k}l_{k+1}\cdot\cdot\cdot l_{D}}
\end{equation}
The fact that these are $Ad(G)$-invariant polynomials, $f_{k}(gXg^{-1})=f_{k}(X)$, is immediate since $det(gXg^{-1})=det(X)$. Also we have that $det(\lambda+X)=det(\lambda-X)$, since for $X\in \mathcal{G}:X^{T}=-\eta X \eta$. Noting that $f_{k}(-X)=(-1)^{k}f_{k}(X)$, we find that necessarily $f_{2k+1}=0$ for $k=0,1,\cdot\cdot\cdot$. For $k=D, f_{D}(X)=det(X)$, while for $k<D$, $f_{k}(X)$ is the dimensional continuation to dimension $D$ of the determinant expression in dimension $k$. In dimension $k$ we have
\begin{equation}
 det(X)=-det(\bar{X}),\;\;\bar{X}=X\eta \in so(k)
 \end{equation}
For $k$ even this yields
\begin{equation}
det(X)=-\left(\frac{1}{(\frac{k}{2})!2^{k/2}}\right)^{2}\varepsilon_{i_{1}\cdot\cdot\cdot i_{k}}\varepsilon_{j_{1}\cdot\cdot\cdot j_{k}}X^{i_{1}i_{2}}\cdot\,\cdot\cdot\cdot\,\cdot X^{i_{k-1}i_{k}}X^{j_{1}j_{2}}\cdot\,\cdot\cdot\cdot\,\cdot X^{j_{k-1}j_{k}},
\end{equation}
where $X^{ij}\equiv X^{i}_{k}\eta^{kj}\, (i,j=0,1,...\, ,k-1)$. For $k$ odd, $det(X)=0$. In dimension $D=2m$ eq.(11) is therefore identical with
\begin{equation}
f_{k}(X)=-\left(\frac{1}{(\frac{k}{2})!2^{k/2}}\right)^{2}\eta_{i_{1}\cdot\cdot\cdot i_{k}j_{1}\cdot\cdot\cdot j_{k} }X^{i_{1}i_{2}}\cdot\,\cdot\cdot\cdot\,\cdot X^{i_{k-1}i_{k}}X^{j_{1}j_{2}}\cdot\,\cdot\cdot\cdot\,\cdot X^{j_{k-1}j_{k}},
\end{equation}
\begin{equation}
\eta_{i_{1}\cdot\cdot\cdot i_{k}j_{1}\cdot\cdot\cdot j_{k} }\equiv \sum_{\sigma\in S_{k}}\varepsilon (\sigma)\eta_{i_{1}j\sigma(1)}\cdot\cdot \cdot \eta_{i_{k}j\sigma(k)}=\delta^{k_{1}\cdot\cdot\cdot k_{k}}_{j_{1}\cdot\cdot\cdot j_{k}}\eta_{i_{1}k_{1}}\cdot\cdot\cdot \eta_{i_{k}k_{k}}
\end{equation}
for $k=2l,\, l=0,1,\cdot\cdot\cdot\, ,D/2 $ and where indices run over $\lbrace0,1,\cdot\cdot\cdot\, ,D-1\rbrace$. If $k=D$, we have in particular: $f_{D}(X)=-(h_{m}(X))^{2}$, where,
\begin{equation}
h_{m}(X)= \frac{1}{m!2^{m}}\varepsilon_{j_{1}\cdot\cdot\cdot j_{D}}X^{j_{1}j_{2}}\cdot\,\cdot\cdot\cdot\,\cdot X^{j_{D-1}j_{D}}
\end{equation}
is the Pfaffian of $X$. Let $A(g)$ be a $D$-dimensional representation of $G=SO(1,D-1)$. The property $A(g^{-1})^{i}_{j}=\eta^{ik}A(g)^{l}_{k}\eta_{lj}$ implies that for $X\in \mathcal{G}=so(1,D-1)$
\begin{equation}
(Ad(g)X)^{ij}=Ad(g)^{i}_{k}X^{k}_{l}A(g^{-1})^{l}_{m}\eta^{mj}=A(g)^{i}_{k}A(g)^{j}_{m}X^{km}
\end{equation}
Since $\varepsilon_{i_{1}\cdot\cdot\cdot i_{D}}$ is an $SO(1,D-1)$-tensor, it is then easily verified that $h_{m}(Ad(g)X)=h_{m}(X)$, i.e., the polynomial function $h_{m}$ is $Ad(G)$-invariant.\\
Define for $D=2m)$
\begin{equation}
l_{m}\equiv h_{m}+f_{m}
\end{equation}
then  $l_{m}\in P_{G}\medskip^{m}(\mathcal{G})$. Remark that $f_{m}=0$ if $m\neq 2k, k=0,1,\cdot\cdot\cdot$. It is the Weil polynomial  $L_{m}\in S_{G}\medskip^{m}(\mathcal{G})$, corresponding to $l_{m}$ under the isomorphism $\mathcal{S}$ (eq.(2)), that we will use in the construction of the Lagrangian form on $P$ of the type given by eq.(6). With eq.(2) we obtain that for $X_{1},\cdot\cdot\cdot ,\, X_{m}\in \mathcal{G}$:
\begin{multline}
L_{m}(X_{1},\cdot\cdot\cdot ,\, X_{m})\equiv (\mathcal{S}l_{m})(X_{1},\cdot\cdot\cdot ,\, X_{m})\\=(C_{1}\varepsilon_{j_{1}\cdot\cdot\cdot j_{D}}+C_{2}\eta_{j_{1}\cdot\cdot\cdot j_{m},j_{m+1}\cdot\cdot\cdot j_{D} })X_{1}\medskip^{j_{1}j_{2}}\cdot\, \cdot\cdot\cdot\,\cdot X_{m}\medskip^{j_{D-1}j_{D}}
\end{multline}
If $m=2k,\, k=0,1,\cdot\cdot\cdot$ , then $C_{1}=(m!2^{m})^{-1}$, $C_{2}=((m/2)!2^{m/2})^{-2}$. For $m\neq 2k, k=0,1,\cdot\cdot\cdot\,$ , $C_{2}=0$.\\
Equation (8) now yields
\begin{multline}
L_{m}(J_{i_{1}i_{2}}, \cdot\cdot\cdot\, ,J_{i_{D-1}i_{D}})=(C_{1}\varepsilon_{j_{1}\cdot\cdot\cdot j_{D}}+C_{2}\eta_{j_{1}\cdot\cdot\cdot j_{m},j_{m+1}\cdot\cdot\cdot j_{D} })\\ \quad\quad\quad\quad\quad\;\times(\delta_{i_{1}}^{j_{1}}\delta_{i_{2}}^{j_{2}}-\delta_{i_{2}}^{j_{1}}\delta_{i_{1}}^{j_{2}})\cdot\,\cdot\cdot\cdot\,\cdot(\delta_{i_{D-1}}^{j_{D-1}}\delta_{i_{D}}^{j_{D}}-\delta_{i_{D}}^{j_{D-1}}\delta_{i_{D-1}}^{j_{D}})\\
=2^{m}(C_{1}\varepsilon_{i_{1}\cdot\cdot\cdot i_{D}}+C_{2}\eta_{i_{1}\cdot\cdot\cdot i_{m},i_{m+1}\cdot\cdot\cdot i
_{D} })\quad\quad\quad\quad\quad\quad\quad\quad\quad\quad \;
\end{multline}
where we made use of the antisymmetry of $\varepsilon_{j_{1}\cdot\cdot\cdot j_{D}}$. From eqs. (4), (9) and (21) we now find for the $L$-form on $P$ defined in eq.(6)
\begin{multline}
L_{m}(\Omega (\omega)+\alpha h(\theta))=(\Omega^{i_{1}i_{2}}\wedge\cdot\cdot\cdot\wedge\Omega^{i_{D-1}i_{D}}+m\alpha \Omega^{i_{1}i_{2}}\wedge\cdot\cdot\cdot\wedge\Omega^{i_{D-3}i_{D-2}}\wedge\theta^{i_{D-1}}\wedge\theta^{i_{D}}\\
+\cdot\cdot\cdot + m\alpha^{m-1}\Omega^{i_{1}i_{2}}\wedge\theta^{i_{3}}\wedge\cdot\cdot\cdot\wedge\theta^{i_{D}}+\alpha^{m}\theta^{i_{1}}\wedge\cdot\cdot\cdot\wedge\theta^{i_{D}})(C_{1}\varepsilon_{i_{1}\cdot\cdot\cdot i_{D}}+C_{2}\eta_{i_{1}\cdot\cdot\cdot i_{m},i_{m+1}\cdot\cdot\cdot i
_{D} })
\end{multline}
Making use of the Bianchi identity for zero torsion: $\Omega^{ij}\wedge\theta_{j}=0$ and of $\theta^{i}\wedge\theta^{j}\eta_{ij}=0$, we obtain finally 
\begin{multline}
L_{m}(\Omega (\omega)+\alpha h(\theta))=\sum_{p=0}^{m-1}\alpha_{p}\varepsilon_{i_{1}\cdot\cdot\cdot i_{D}}\Omega^{i_{1}i_{2}}\wedge\cdot\cdot\cdot\wedge\Omega^{i_{2p-1}i_{2p}}\wedge\theta^{i_{2p+1}}\wedge\cdot\cdot\cdot\wedge\theta^{i_{D}}\\+H_{m}(\Omega)+F_{m}(\Omega)\quad\quad\quad\quad\quad\quad\quad\quad\quad\quad\quad\quad\quad\quad\quad\quad\quad\;\; 
\end{multline}
with
\begin{equation}
H_{m}(\Omega)=C_{1}\varepsilon_{i_{1}\cdot\cdot\cdot i_{D}}\Omega^{i_{1}i_{2}}\wedge\cdot\cdot\cdot\wedge\Omega^{i_{D-1}i_{D}}
\end{equation}
\begin{equation}
F_{m}(\Omega)=C_{2}\eta_{i_{1}\cdot\cdot\cdot i_{m},i_{m+1}\cdot\cdot\cdot i_{D} }\Omega^{i_{1}i_{2}}\wedge\cdot\cdot\cdot\wedge\Omega^{i_{D-1}i_{D}}
\end{equation}
and where $F_{m}(\Omega)\neq 0$ only if $m=2k$.\\
The $L$-form (23) on $P$ projects to a unique $D$-form $\bar{L}_{m}(\Omega (\omega)+\alpha h(\theta))$ on $M$. The first term in eq. (23) gives the Lovelock Lagrangian for even dimension $D$. Remark that the coefficients
\begin{equation}
\alpha_{p}=C_{1}\left( \begin{matrix}
 m \\ 
 m-p
 \end{matrix}\right)
 \alpha^{m-p}
\end{equation}
are not completely arbitrary as in the original Lovelock Lagrangian. Here $\alpha$ is a (fundamental) constant of dimension $(length)^{-2}$. The projection of the second and third terms $\bar{H}_{m}(\Omega)$ and $\bar{F}_{m}(\Omega)$ are closed $D$-forms on $M$. They give topological invariant integrals on $M$ and can be neglected in the classical but not in the quantum domain when there are quantum fluctuations in the topology of spacetime. Their de Rham cohomology classes $[\bar{H}_{m}(\Omega)]$ and $[\bar{F}_{m}(\Omega])$ are (up to a constant in eq.(24) and (25)) respectively the Euler class and the last Pontrjagin class.[6,7]

\newpage 
\begin{center}
\textbf{\Large References}
\end{center}
$[1]$ Lovelock D, J.Math.Phys. \textbf{12} (1971), 498.\\
$[2]$ Müller-Hoissen F, Class.Quantum Grav. \textbf{3} (1986), 665.\\
$[3]$ Zumino B, Phys. Rep. \textbf{137} (1986), 109.\\
$[4]$ Teitelboom C and Zanelli J, Class Quantum Grav. \textbf{4} (1987), L124.\\
$[5]$ Kakazu K and Matsumoto S, Prog. Theor. Phys. \textbf{78} (1987), 166.  \\
$[6]$ Kobayashi S and Nomizu K, Foundations of Differential Geometry

 (Wiley, New York,1963), vols 1 and 2.\\
$[7]$ Spivak M, A Comprehensive Introduction to Differential geometry (Publish or Perish,

 Berkeley, 1979), vols 1-5.\\
$[8]$ Bleecker D, Gauge Theory and Variational Principles (Addison-Wesley, 1981).\\
$[9]$ Vaisman I, Cohomology and Differential Forms (Marcel Dekker, New York, 1973).\\
$[10]$ Greub W, Halperin S and Vanstone R, Connections  Curvature and Cohomology

 (Academic Press, New York, 1973), vols. 1 and 2. \\

\end{document}